# Tunable high Chern-number quantum anomalous Hall effect through interlayer ferromagnetic coupling in two-dimensional ferromagnet NiSbO$_3$

Xuebing Peng[†], Mingsu Si,[‡] Daqiang Gao[†]

[†]School of Physical Science and Technology, Lanzhou University, Lanzhou 730000, China

[‡]School of Materials and Energy, Lanzhou University, Lanzhou 730000, China

**ABSTRACT:** The high Chern-number quantum anomalous Hall effect (QAHE) is significant and fascinating due to the presence of multiple dissipationless chiral edge states. Here, we predict that monolayer NiSbO$_3$ possesses the Chern number C = 3, confirmed by the anomalous Hall conductance and the chiral edge states. The magnetic anisotropic energy (MAE) responsible for ferromagnetic order is 0.641 meV originating from Ni-*d* and Sb-*p* orbitals, where the contributed MAE from same spin-up channels predominates. In forward electric fields, the negative MAE makes the easy magnetization direction perpendicular to the surface, which is conducive to the realizing of high Chern-number QAHE. The simulated Curie temperature is 291 K. Intriguingly, in a bilayer, the obtained C = 6 is twice that of the monolayer, thanking to the interlayer ferromagnetic coupling. Our work offers a promising candidate for potential applications in topological quantum devices and spintronics.

As one of the crucial members of Hall family, quantum anomalous Hall effect (QAHE) has garnered significant attention in condensed matter physics.[1-3] Distinct from quantum Hall effect, QAHE exhibits a quantized conductance under zero magnetic flux, which is proposed by Haldane in theory.[4] In experiments, it directly leads to the anomalous Hall conductance (AHC). Due to the topologically protected edge states, QAHE offers a dissipationless electron transport, which has potential applications in the next-generation low-power-consumption electronic devices, such as efficient quantum bits, quantum sensors and quantum gate operations.[5-7] It is also an idea platform to study topological superconductivity and Majorana fermions.[8-10] QAHE requires the breaking of time-reversal symmetry. Three methods are usually used: doping magnetic elements, magnetic proximity effect and intrinsic magnetic topological insulator. In 2013, Chang et. al. successfully doped Cr atom into (Bi, Sb)$_2$Te$_3$, confirming QAHE in experiment. However, the magnetic inhomogeneity severely limits observation of temperature.[11] In the proximity coupling of EuS/Bi$_2$Se$_3$, one is difficult to control charge transfer concentrations, putting the chemical potential inside bulk conduction band, Hence, topological surface states may not have involvement.[12, 13] Later, in low-dimensional ferromagnetic MnBi$_2$Te$_4$, one observed QAHE once again.[14] While some 2D ferromagnetic materials that exhibit QAHE have been theoretically predicted.[15-18] But their low Chern-number $C = 1$ may not sufficiently resist the contact resistance between a metal electrode and Chern insulator in the envisioned interconnect devices. If one can seek a high-Chern-number QAHE, it is imperative since it can offer more edge states without dissipation to increase the conductance of electron.

How to realize QAHE with the high Chern number? one way is to increase the doping concentration, but presenting the inhomogeneity harming the performance of device.[19, 20] Another one is to search for a high-symmetry atomic structure, where the interlayer magnetic atoms in a bilayer are coupled by ferromagnetic (FM) states. As is well known, when considering the spin orbital coupling (SOC) in the spin-polarized band dispersions with linearly crossing points, orbital inverting can result in the emergence of nontrivial topology.[21-23] If such atomic structure exits a very high symmetry, then more crossing points appear within the first Brillouin zone and contribute the high Chern number. While the interlayer FM coupling can retain QAHE in multiple layers, and provide more Chern numbers. This is the case of multilayer MnBi$_2$Te$_4$. [24] But insufficiently, one needs to obtain a forced FM state in the even layers with antiferromagnetic (AFM) states. Thus, to realize the high-Chern-number QAHE, our work seeks a high-symmetry 2D ferromagnetic material, and magnetic atoms between the bilayer are coupled by a FM state. Furthermore, early Mermin−Wagner theorem thought the magnetism does not exists in the 2D materials due to thermal disturbance.[25] However, the existence of sizable the magnetic anisotropic energy (MAE) can resist this disturbance, and stabilize FM long-range order in low-dimensional systems. In addition to this, MAE is capable of responsible for the easy magnetization direction, and can be tuned by a strain and an external electric field.[26-28] Hence, investigating MAE in topological materials is valuable for the scientific research and practical applications.

In this work, electronic structure, topological properties, and MAE of 2D NiSbO$_3$ are systemically investigated via the combination of DFT+$U_{eff}$ and tight-bonding model. Monolayer NiSbO$_3$ possesses a ferromagnetic state, and the magnetic moment per Ni$^{3+}$ ion is 1 $\mu_B$. Weyl point appears around Fermi level. There exists six Weyl points within the first Brillouin zone owing to the $C_{3h}$ rotation and the inversion symmetry. Under SOC, Weyl points disappear and are gaped by 16.6 meV when establishing out-of-plane magnetic moments. The obtained QAHE has the Chern number $C = 3$. While for in-plane magnetic moments, Weyl points are retained. In the bilayer with AA stacking manner, NiSbO$_3$ is also ferromagnetic. Interestingly, the obtained Chern number $C = 6$. In addition, we explore the origin of MAE by atom-, spin- and orbital-decomposition methods for the monolayer. Upon using forward electric fields, the sign of MAE can be changed. The obtained out-of-plane easy magnetization helps to the realization of QAHE. Finally, we simulate the Curie temperature, and the estimated $T_c$ is 291 K. Our findings provide a desired candidate toward the topological quantum devices with multiple chiral edge states.

We employ first-principles calculations to investigate 2D NiSbO$_3$ based on the density functional theory (DFT) of the projector augmented wave (PAW) approach, as implemented in the Vienna Ab initio Simulation Package (VASP).[29, 30] The Perdew−Burke−Ernzerhof (PBE) functional of generalized gradient approximation (GGA) is utilized to solve the exchange correlation interaction between electrons.[31] The Hellmann-Feynman forces on each atom are smaller than 0.001 eV/Å during structural relaxation. To obtain accurate results, the convergence criterion of total energy and a plane-wave cutoff energy are set to 10$^{-7}$ eV and 520 eV. We sample a 9×9×1 Monkhorst-Pack $k$-point mesh within the first Brillouin zone for self-consistent calculations, and set a denser 20 × 20 × 1 $k$-point mesh to computed MAE. To simulate 2D system, a vacuum region of 20 Å along the direction perpendicular to the surface is added to eliminate the interaction between the periodic images. A long-range van der Waals interaction for bilayer NiSbO$_3$ is considered via DFT-D3 method.[32] The correlation effects for Ni-$d$ localized orbitals are solved via applying an effective on-site Hubbard correction $U_{\text{eff}}$ = 3 eV.[23, 33] The maximally localized Wannier function (MLWF) tight-binding model for Sb-$p$, O-$p$, and Ni-$d$ orbitals is used to compute the Berry curvature and AHC as implemented in WANNIER90 software package.[34, 35] WANNIERTOOLS package is taken to computes the edge states and the Wannier charge centers.[36] Phonon dispersions are obtained by using phonon code.[37] Ab initio molecular dynamics simulations are performed to evaluate the thermodynamic stability. Post-processing of data is carried out using VASPKIT code.[38] The Curie temperature is estimated by mcsolver code.[39]

**Results and discussion.** The monolayer NiSbO$_3$ crystallizes as the hexagonal lattice with the space group of $P31m$. One conventional unit cell contains five atomic layers with Sb–O–Ni–O–Sb, as shown in Fig. 1(a). The optimized lattice constants are a = b = 5.00 Å. The magnetic moment per Ni atom is 1 $\mu_B$. The Ni-$d$ orbital inside O octahedron is splitting due to an asymmetric crystal field. Originally degenerate five $d$-resolved orbitals are splitting into three sets of orbitals a$_1$ ($d_{z^2}$), e$_1$ ($d_{xz}/d_{yz}$) and e$_2$ ($d_{x^2-y^2}/d_{xy}$) from the partial density of states (PDOS) (see Fig. S1(a) of supplemental material (SM)). $d_{xz}$ and $d_{yz}$ as well as $d_{x^2-y^2}$ and $d_{xy}$ are degenerate under the rotating symmetry $C_{3h}$ in point group $D3d$. Since this system possesses magnetism, the spin channel in orbital is splitting, as shown in Fig. 1(b). In the oxidation state, that is Ni$^{3+}$ ion with seven electrons, six electrons occupy the a$_{1,\uparrow\downarrow}$ and e$_{2,\uparrow\downarrow}$ orbitals because the crystal field is larger than the exchange field. The remaining one electron occupies the up-spin e$_1$ orbital, called the quarter-filled state.[40] Through the integral of e$_{1\uparrow}$ orbital on PDOS, the obtained electrons are 1.16$e$ in line with the result of our analysis. In addition, we check the structural stability from both phonon spectrum and molecular dynamics (see Fig. S1(b) and S1(c) of SM). The computed phonon spectrum indicates no imagery frequency, guaranteeing the high dynamic stability. Molecular dynamic simulations show total energy variation with a tiny range oscillation and no broken atomic structure during the time duration round 7 ps.

To determine magnetic ground state for the monolayer NiSbO$_3$, we compute the exchange energy through constructing one FM and three AFM configurations (see Fig. S2 of SM), which is defined as the difference between the total energies of AFM and

FM configurations, or $\Delta E_{\text{exc}} = E_{\text{AFM}} - E_{\text{FM}}$. Our obtained $\Delta E_{\text{exc}}$ is positive, illustrating the monolayer NiSbO$_3$ of a FM ground state. This FM ground state coincides with the Goodenough-Kanamori-Anderson rules.[41-43] The nearest-neighbor two Ni atoms have indirect and direct exchange interactions. The former is mediated by the intermediate O atoms of FM coupling, whereas the latter is referred to the AFM coupling. Due to the localization of $d$ orbital in each Ni atom, the AFM coupling is usually weaker than the FM coupling. It is also noting that Ni–O–Ni bond angle of 91° is close to 90°, usually providing the FM state in superexchange interaction.

The spin-polarized band structures are plotted in Figs. 1(c) and 1(d). The band structure of spin-down channel along Γ–M–K–Γ indicates an insulating state with the band gap of 2.13 eV. Whereas that of spin-up channel is gapless near Fermi level, existing in linearly crossing point along Γ–M. Distinct from the Dirac point, this crossing point is a doubly degenerate Weyl point. Such band character is a typical Weyl semimetal. For the spin-up channel, the bands near the Fermi level are mainly occupied by the Ni-$e_1$ orbital. When considering SOC, the Weyl point is not stable. Nonzero matrix element $\langle d_{yz}|\hat{L}_z|d_{xz}\rangle = i$ with $\hat{L}_z$ being the $z$ component of orbital angular momentum operator $\hat{L}$ has the shift on energy (see Table I), opening a gap of around 16.6 meV, as shown in Fig. 1(e). This is the origin of the nontrivial topological state. Considering $C_{3h}$ and the inversion symmetry, six opening gaps would appear within the first Brillouin zone, as shown in Fig. 1(f).

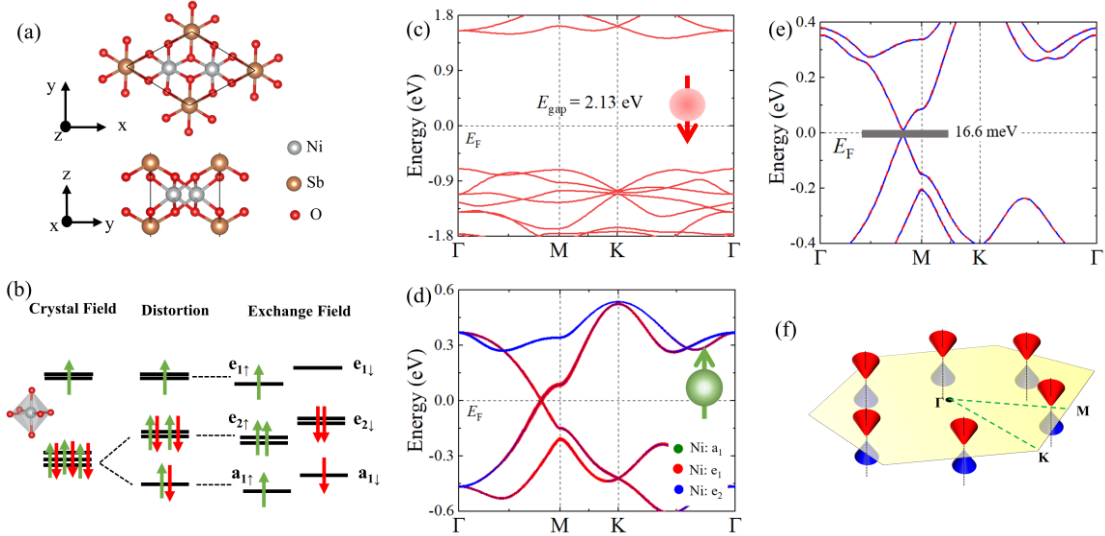

FIG. 1. (a) Top (top panel) and side (bottom panel) views of atomic structure in monolayer NiSbO$_3$. (b) Diagram of Ni-$d$ splitting orbital from the crystal field of the oxygen octahedron. The $a_1$, $e_1$ and $e_2$ symbols are $d_{z^2}$, $d_{xz}/d_{yz}$ and $d_{x^2-y^2}/d_{xy}$ orbitals, respectively. The ↑↓ denotes the spin direction. Band structures of (c) spin-down and spin-up channels (d). (e) The band structure with spin orbital coupling (SOC). Blue solid and red dotted lines represent the obtained band structures of DFT scheme and Wannier90 fitting, respectively. (f) Three-dimension diagram of band dispersions around Fermi level within the first Brillouin zone. The Fermi level is set to zero.

To explore the nontrivial topological state in monolayer NiSbO$_3$, we firstly use MLWF in the form of tight-bonding model to fit the band structure near the Fermi level.

The obtained band structures are consistent with those computed by DFT, as shown in Fig. 1(e). The Berry curvatures are calculated by the Kubo formula[44]

$$\Omega_z(\boldsymbol{k}) = -2 \sum_n \sum_{m \neq n} f_n(\boldsymbol{K}) \frac{\hbar^2 \text{Im}\langle \psi_{n\mathbf{k}}|\hat{v}_x|\psi_{m\mathbf{k}}\rangle\langle \psi_{m\mathbf{k}}|\hat{v}_y|\psi_{n\mathbf{k}}\rangle}{(E_{m\mathbf{k}} - E_{n\mathbf{k}})^2}, \quad (1)$$

where $f_n$, $\hbar$, $\hat{v}_{x/y}$, and $E_{i\mathbf{k}}$ denote Fermi Dirac distribution function, the reduced Planck constant, the velocity operator along $x/y$ direction, and the eigenvalue of Bloch wavefunction $\psi_{i\mathbf{k}}$, respectively. $\Omega_z(\boldsymbol{k})$ is obtained by the interpolated tight-binding Hamiltonian. Figure 2(a) shows the localization of Berry curvatures, which mainly originates from the $e_{1\uparrow}$ orbital around the Fermi level. The inset displays the distribution of Berry curvatures within the first Brillouin zone, where the Berry curvatures have the same signs and appear at $M_1$, $M_2$, and $M_3$.

By integrating $\Omega^z(\mathbf{k})$ on the entire first Brillouin zone, AHC[45] is obtained as

$$\sigma_{xy} = \frac{e^2}{h} \int_{BZ} \frac{\Omega^z(\mathbf{k})}{2\pi} d^2\mathbf{k}, \quad (2)$$

where we use $e^2/h$ as the units of $\sigma_{xy}$. AHC can be also written as $\sigma_{xy} = Ce^2/h$ with the coefficient $C$ being the Chern number.[46] Figure 2(b) exhibits a wide platform of proximately 17 meV. $\sigma_{xy} = 3e^2/h$ is obtained for the monolayer NiSbO$_3$. This means that monolayer NiSbO$_3$ is a high Chern-number 2D ferromagnet. Since each localized Berry curvature contributes $C = 1/2$, the six ones under the $C_{3h}$ rotation symmetry gives rise to $C = 3$. This is also confirmed by Wannier charge centers (WCCs), which is related to the Chern number $C = \frac{1}{a}\sum_n[\bar{x}_n(k_y) - \bar{x}_n(0)]$,[47] with $a$ and $\bar{x}_n(k_y)$ being the lattice constant and a smooth function of $k_y$ for $k_y \in [0,2\pi]$, respectively. Here the Chern number can be viewed as the number of pumped electronic charges across one unit cell in a course of cycle. WCCs give the winding number 3, which is consistent with the Chern number $C = 3$, as shown in the inset of Fig. 2(b). This high Chern number 2D ferromagnet is rare in comparison with the reported topology materials such as V$_2$O$_3$,[48] Mn$_2$C$_{18}$H$_{12}$,[15] V$_2$MX$_4$[17] that exhibit $C = 1$. The high-Chern-number QAHE is capable to provide more chiral edge channels for dissipationless electronic transport.

Our computed topological edge states based on the iterative Green's function method are displayed in Fig. 2(c). Three gapless chiral edge states between valence band maxima and conduction band minima appear in the bulk bands. In experiments, one can detect this quantum state via noncontact magneto-optical techniques. The electronic-transport diagram of high Chern number in QAHE is illustrated in Fig. 2(d). Three spin-up channels are capable of orderly transmitting electrons to avoid the scatter between them. This system can be used to build next-generation low-power-consumption electronic devices. If the direction of magnetic moment is reversed to an opposite direction, the positive Chern number is changed to $C = -3$ (see Fig. S3 of SM), suggesting that the electronic-transport direction can be switched in the chiral edge channels. When we tune the magnetic moment to be in-plane aligned, the Weyl point remains (see Fig. S4 of SM).

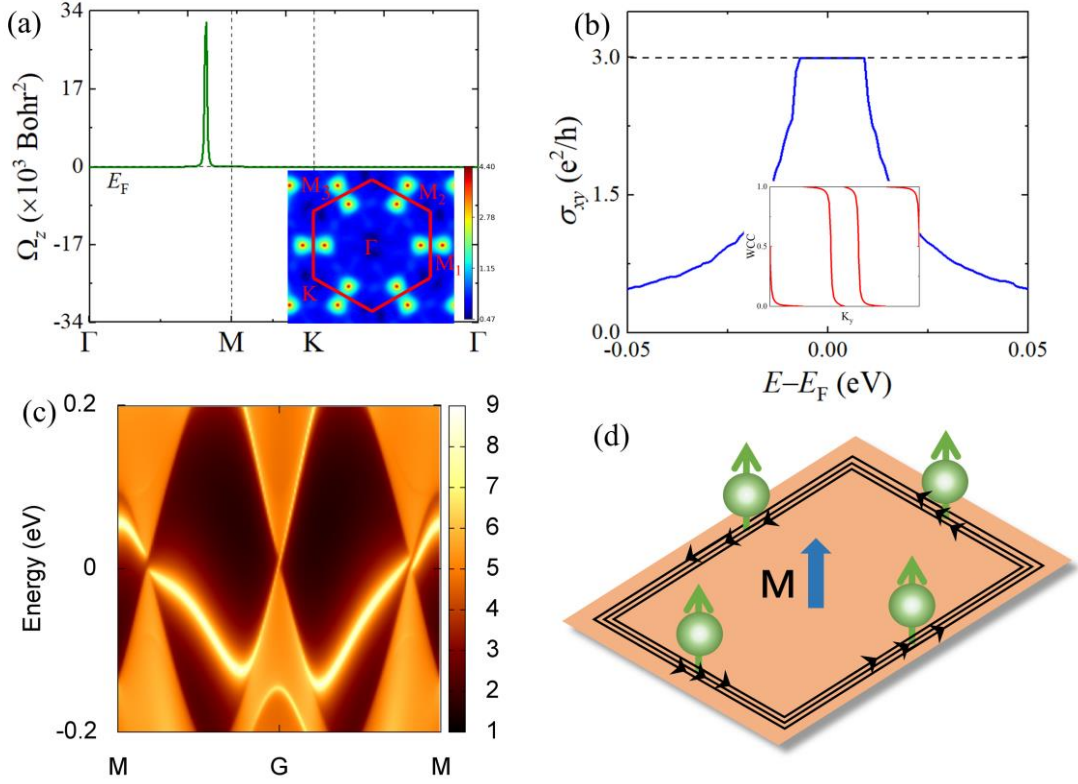

FIG. 2. (a) Berry curvatures along high-symmetry paths. The inset is the berry curvatures within the first Brillouin zone. (b) Anomalous Hall conductance (AHC) of chemical potentials from –0.05 to 0.05 eV. The inset displays Wannier charge centers (WCCs). (c) Topological edge spectrum. (d) Diagram of electronic transport, where M is the magnetization direction. The Fermi level is set to zero.

To explore the QAHE of layer effect, we construct bilayer $NiSbO_3$ along $z$ direction. There usually exist AA and AB stacking manners (see Fig. S5 of SM). To determine their magnetic ground states, we set the FM and AFM configurations. The former has the same spin directions in interlayer magnetic atoms, whereas the latter has the opposite spin directions. Compared to the AFM configuration, FM configuration exhibits a lower energy for the AA and AB stackings, indicating the bilayer $NiSbO_3$ favors a FM ground state. This is different from bilayer $MnBi_2Te_4$[49] and $CrI_3$[50] that exhibit an AFM ground state. Figure 3(a) shows the spin-polarized band structure in the AA stacking manner. The band structure is similar to that of monolayer. The linearly crossing point reserves around the Fermi level. Upon considering SOC, the linearly crossing point disappears, opening a gap of proximately 5.1 meV, as shown in Fig. 3(b). To verify the existence of QAHE, we compute AHC. Interestingly, a platform appears around the Fermi level, and $\sigma_{xy} = Ce^2/h = 6\,e^2/h$, as shown in Fig. 3(c). This corroborates that QAHE can exist in the bilayer $NiSbO_3$. However, for the bilayer $NiSbO_3$ in the AB stacking manner, no linearly crossing point appears in the spin-polarized band structure (see Fig. S5(c) of SM). It is suggested that the bilayer $NiSbO_3$ with AB stacking may not support QAHE.

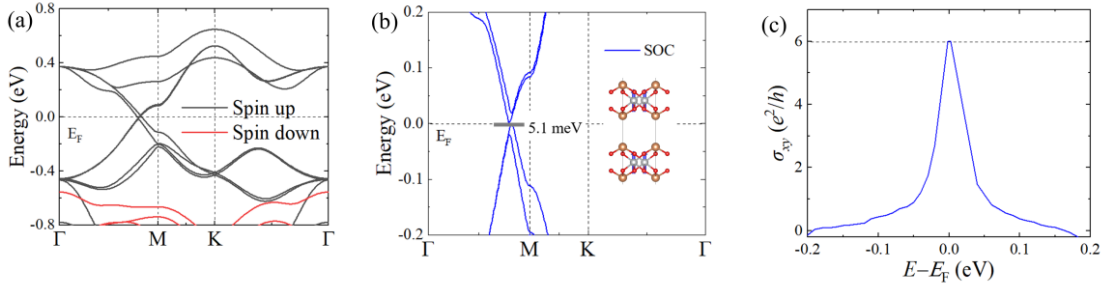

FIG. 3. (a) Spin polarized band structure of bilayer $NiSbO_3$. Black and red lines are spin-up and spin-down channels. (b) Band dispersions upon considering SOC. The inset shows the atomic structure of bilayer $NiSbO_3$ in AA stack. (c) AHC of chemical potentials from –0.2 to 0.2 eV. The Fermi level is set to zero.

MAE is capable of stabilizing long-range FM order and determining the direction of easy magnetization in 2D materials. This forces us to explore it, defined as $E_{MAE} = E(\perp) - E(\parallel)$ with $E(\perp)$ and $E(\parallel)$ being the energies of magnetic moments along out-of-plane ($\perp$) and in-plane ($\parallel$) directions, respectively. The computed MAE is 0.641 meV/f. u., which is comparable to those of preparing 2D ferromagnets experimentally such as $CrI_3$ (0.803 meV/f. u.),[51] $CrGeTe_3$ (0.66 meV/f. u.),[52] and $Fe_5GeTe_2$ (0.58 meV/f. u.).[53] The positive MAE shows the in-plane easy magnetization. Here, we discuss the origin of MAE. MAE is estimated by $E_{MAE} \approx \frac{1}{2}\sum_i \Delta E_{SOC}^i$,[54] where $\Delta E_{SOC}^i$ denotes difference between $E(\perp)$ and $E(\parallel)$ in the $i$th atom under SOC. The obtained MAE from O atom gets close to zero due to weak SOC effect. MAE originates from Ni and Sb atoms. They contribute 0.306 and 0.341 meV, respectively. Essentially, MAE arises from the coupling strength between occupied ($o$) and unoccupied ($u$) states via the orbital angular momentum operator. MAE in SOC second order perturbation is written as spin-conversed (SC) and spin-flit (SF) terms[55]

$$E_{MAE}^{total} = E_{MAE}^{SC} + E_{MAE}^{SF}, \qquad (3)$$

for the right first term

$$E_{MAE}^{SC} = \xi^2 \sum_{o,u} \frac{\left|\langle o^{\downarrow(\uparrow)}|\hat{L}_x|u^{\downarrow(\uparrow)}\rangle\right|^2 - \left|\langle o^{\downarrow(\uparrow)}|\hat{L}_z|u^{\downarrow(\uparrow)}\rangle\right|^2}{E_u^{\downarrow(\uparrow)} - E_o^{\downarrow(\uparrow)}} \qquad (4)$$

and another term

$$E_{MAE}^{SF} = \xi^2 \sum_{o,u} \frac{\left|\langle o^{\uparrow(\downarrow)}|\hat{L}_z|u^{\downarrow(\uparrow)}\rangle\right|^2 - \left|\langle o^{\uparrow(\downarrow)}|\hat{L}_x|u^{\downarrow(\uparrow)}\rangle\right|^2}{E_u^{\downarrow(\uparrow)} - E_o^{\uparrow(\downarrow)}}, \qquad (5)$$

where $\hat{L}_{z(x)}$ and $E_{o,u}^{\downarrow(\uparrow)}$ denote the $z(x)$ component of orbital angular momentum operator, the energies of occupied or unoccupied state with spin direction $\downarrow(\uparrow)$, respectively. In denominator, $E_o \leq E_F$ and $E_u > E_F$. The sign of MAE depends on angular momentum matrices in numerator. In spin-conversed term, the obtained MAE is 3.841 meV responsible for the in-plane easy magnetization, of which the same spin-up channels contribute 3.41 meV. Whereas the spin-flit MAE is –3.216 meV responsible for the out-of-plane easy magnetization. For the total MAE, the coupling between same spin-up channels predominates, which is related of the distribution of spin channel near the Fermi level.

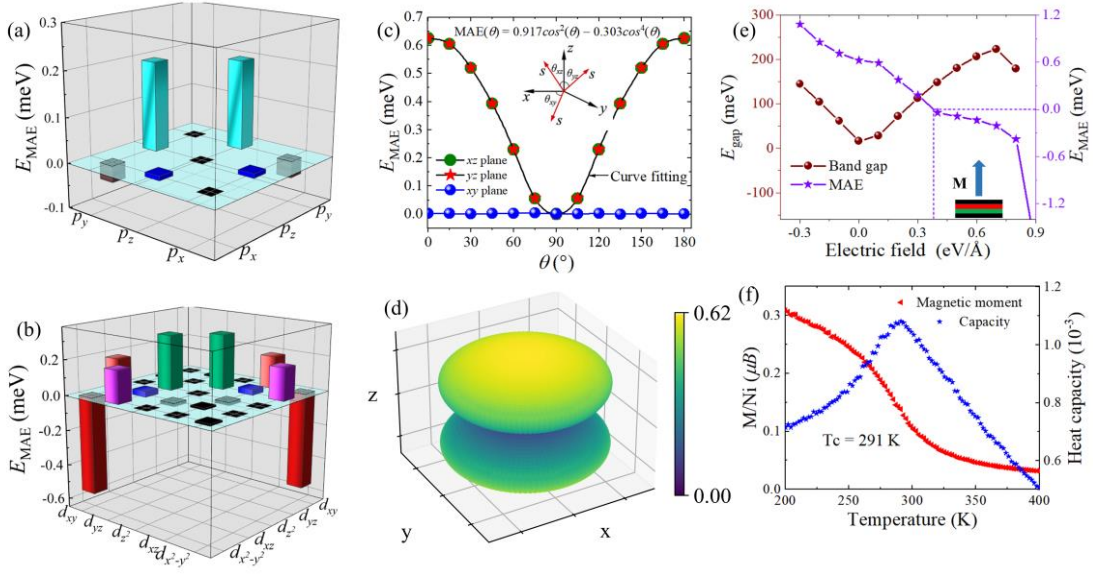

FIG. 4. The magnetic anisotropy energies (MAEs) of (a) $p$-resolved and (b) $d$-resolved orbitals. (c) MAE variation vs $\theta_{xy}$, $\theta_{xz}$ and $\theta_{yz}$. The inset is angle $\theta$ in Cartesian coordinate. (d) Three-dimension spatial distribution of MAE. (e) Tunable bandgap and MAE through electric field. The arrow inside dashed box denotes the emergence of out-of-plane easy magnetization. (f) Magnetic moment and heat capacity vs temperature.

Next, we project MAE onto $p$- and $d$-resolved orbitals. The matrices of $\hat{L}_{z/x}$ between different $p$ ($d$) orbitals are listed in Table I. The contributed MAE in the $p$ orbital is almost the same as that of $d$ orbital, with being 0.310 and 0.345 meV, respectively. Majority of coupling states provide the positive MAE shown in Fig. 4(a) and 4(b), favoring the in-plane easy magnetization. Whereas only minority of coupling states are the negative MAE including $(d_{x^2-y^2}, d_{xy})$ and $(p_x, p_y)$, supporting the out-of-plane easy magnetization. The absolve value of coupling state $(d_{x^2-y^2}, d_{xy})$ is the largest, as $\langle d_{xy}|\hat{L}_z|d_{x^2-y^2}\rangle = 2i$ is the largest matrix element. Furthermore, we explore the angular-dependent MAE, denoted as $E_{\mathrm{MAE}} = E(\theta_{xz/yz}, \theta_{xy}) - E(\theta_{xz/yz} = 90°, \theta_{xy} = 0°)$, where $\theta_{xz/yz}$ and $\theta_{xy}$ are the polar and azimuthal angles, as shown in Fig. 4(c). In $xy$ plane, MAE not depends on $\theta_{xy}$. For $xz$ and $yz$ planes, the polar-angular-dependent MAE appears and exhibits a mirror symmetry from 0° to 180°. We use a formula $E_{\mathrm{MAE}}(\theta) = K_1 \cos^2(\theta) + K_2 \cos^4(\theta)$ to fit polar-angular-dependent MAE. The black curve line is the fitted result by the specular angles of MAE, with obtained $K_1$ and $K_2$ being 0.917 and –0.303, respectively. Three-dimension plot of angular-dependent MAE is displayed in Fig. 4(d), where we can directly see the isotropous azimuthal-angular and the anisotropic polar-angular variations.

MAE and bandgap are tunable in the monolayer $NiSbO_3$. Here, we resort to an external electric field perpendicular to the surface. MAE and bandgap have big change, as shown in Fig. 4(e). In forward electric field, the positive MAE is changed to the negative value, illustrating the easy magnetization switched to the out-of-plane direction. The maximum of bandgap can reach up to over 200 meV around 0.75 eV/Å. This is conducive to the realization and detection of QAHE. We also use strain to regulate MAE and bandgap (see Fig. S6 (a) of SM). MAE continues to decline from a compressive to tensile strain, but their values remain positive, making the in-plane easy magnetization unchanged. When applying the tensile strain, the bandgap somewhat increases, whereas it decreases upon the compressive strain.

TABLE I. The matrix elements between different $p$ ($d$) orbitals under orbital angular momentum operator $\hat{L}_\sigma$, where $\sigma = x$ and $z$ are the components of Pauli matrices.

| ⟨o\| | $\hat{L}_\sigma$ ($\sigma = x$ and $z$) | | | | | | | | |
|---|---|---|---|---|---|---|---|---|---|
| \|u⟩ | $p$ orbital | | | | $d$ orbital | | | | |
|  | $p_y$ | $p_z$ | $p_x$ |  | $d_{xy}$ | $d_{yz}$ | $d_{z^2}$ | $d_{xz}$ | $d_{x^2-y^2}$ |
| $p_y$ | 0 | $-i(x)$ | $i(z)$ | $d_{xy}$ | 0 | 0 | 0 | $-i(x)$ | $2i(z)$ |
|  |  |  |  | $d_{yz}$ | 0 | 0 | $-\sqrt{3}i(x)$ | $i(z)$ | $-i(x)$ |
| $p_z$ | $i(x)$ | 0 | 0 | $d_{z^2}$ | 0 | $\sqrt{3}i(x)$ | 0 | 0 | 0 |
|  |  |  |  | $d_{xz}$ | $i(x)$ | $-i(z)$ | 0 | 0 | 0 |
| $p_x$ | $-i(z)$ | 0 | 0 | $d_{x^2-y^2}$ | $-2i(z)$ | $i(x)$ | 0 | 0 | 0 |

Lastly, we estimate the Curie temperature for the monolayer NiSbO$_3$. The Heisenberg Hamiltonian including single-ion anisotropy is expressed as

$$H = -J_{ij} \sum_{i<j} \mathbf{S_i} \cdot \mathbf{S_j} - A \sum_i (S_i^z)^2 , \tag{6}$$

where $J_{ij}$, $\mathbf{S_i}$ and $S_i^z$ are the exchange coupling strength between the $i$th and $j$th magnetic atoms, the spin quantum number of the $i$th Ni atom and the $z$ component of $\mathbf{S_i}$, respectively. $i < j$ aims to avoid the repeating count of $J_{ij}$. Single-ion anisotropy energy $A$ is determined by $E_{\text{MAE}}/|S_i|^2$. In general, considering the exchange couplings of the nearest and the next-nearest neighbors is adequate, signified as ($J_\text{N}$) and ($J_\text{NN}$), respectively. To compute the exchange coupling strength, we set three distinct magnetic configurations, namely one ferromagnet and two antiferromagnets (Néel and zigzag) for a 2×2×1 supercell (see Fig. S2 of SM). $J_\text{N}$ and $J_\text{NN}$ are extracted from

$$E_{\text{AFM1}} - E_{\text{FM}} = 24 J_\text{N} |S|^2 \tag{7}$$
$$E_{\text{AFM2}} - E_{\text{FM}} = 8 J_\text{N} |S|^2 + 32 J_\text{NN} |S|^2.$$

The obtained in-plane exchange coupling strengths are $J_\text{N}(\parallel) = 93.65$ meV and $J_\text{NN}(\parallel) = 37.94$ meV. While the out-of-plane those are $J_\text{N}(\perp) = 92.41$ meV and $J_\text{NN}(\perp) = 37.33$ meV. The remaining term $A = 2.56$ meV. Based on these parameters, using 200×200×1 supercell to avoid the influence of finite size effect during Monte Carlo simulations, the estimated $T_\text{C}$ is proximately 295 K, as shown in Fig. 4(f).

In summary, we systemically investigate the electronic structure, the crystal stability, QAHE and MAE for 2D NiSbO$_3$ via the combination of DFT+$U_{\text{eff}}$ and tight-bonding model. The phonon spectrum and the molecular dynamic simulations corroborate the stability of atomic structure. Its magnetic ground state is ferromagnetic, and the magnetic moment per Ni ion is 1 $\mu_B$. The quarter-filled state of $e_1$ orbital provide the Weyl point. When considering SOC, QAHE appears and the Chern number $C = 3$, determined by AHC and WCCs as well as the topologic edge spectrum. In the bilayer with AA stacking manner, this system is also FM. Particularly, QAHE exists in this stacking, and the Chern number $C = 6$. Through the computed MAE in the monolayer, the easy magnetization is in-plane. MAE stems from Ni-$d$ and Sb-$p$ orbitals. The coupling between same spin-up channels predominates in MAE. In forward electric field, the bandgap increases and the easy magnetization is changed to the out-of-plane direction, which helps to the realization and detection of QAHE. Lastly, the estimated $T_c$ is proximately 291K based on the Monte Carlo simulations.

**ACKNOWLEDGMENTS**

This work was supported by the National Science Foundation of China (Grant No. 11874189), the Natural Science Foundation of Gansu Province, China (Grant No. 20JR10RA648), and Gansu Province Education Science and Technology Innovation Youth Doctor Foundation (2022QB-002). We also acknowledge the Fermi cluster at Lanzhou University for providing computational resources.